# The role of broadband connectivity in achieving Sustainable Development Goals (SDGs).

OGUTU B. OSORO[1] and EDWARD J OUGHTON[1]

[1] Department of Geography & Geoinformation Science Department, George Mason University, Fairfax, VA 22030, USA.

## Abstract

Broadband connectivity is a tool for catalyzing socio-economic development and reducing the societal inequalities. Recent studies have investigated the supporting role of broadband in addressing Sustainable Development Goals (SDGs). Relationally, emerging ultra-dense broadband networks such as 5/6G have been linked to increased power consumption and more carbon footprint. With SDGs recognized as interdependent and addressing one should not jeopardize the achievement of the other, there is need for sustainability research. Despite the need to narrow the digital divide and address the SDGs by 2030, it is surprising that limited comprehensive studies exist on broadband sustainability. To this end, we review 113 peer reviewed journals focusing on six key areas (SDGs addressed, application areas, country income, technology, methodology and spatial focus). We further discuss our findings before making four key recommendations on broadband sustainability research to fast-track SDG achievement by 2030 especially for developing economies.

# I.   Introduction

In 2023, the International Telecommunication Union (ITU) reported that about 2.6 billion people were still unconnected to the Internet. Connecting the unconnected has been an objective of the United Nations (UN) to bring life-changing opportunities and benefits to the population. The persistent gap between individuals with access to Information Communication Technology (ICT) and those with limited or no access (digital divide) threatens the achievement of UN's Sustainable Development Goals (SDGs). Studies have shown that digital divide intensifies inequality between rural and urban residents leading to concepts such as digital poverty [1]. Such inequality jeopardizes the realization of SDG goals and targets that are meant to achieve global prosperity by 2030.

Broadband is a general-purpose technology that fits in supporting socio-economic development and the SDGs. For instance, the technology supports utility and critical infrastructure systems such as roads, electricity, water as well as value addition [2]. The technology can also be used to support and track the progress of the 17 SDGs and the 76 targets [3]. Additionally, the data collection sensors such as drones, satellites and Internet of Things devices provide high spatial-temporal data necessary for applications and projects supporting SDGs.

More importantly, the UN places emphasis on sustainability of all the practices aimed at achieving the SDGs.  While universal broadband coverage is part of UN's SDG 9, specifically Target 9c [4], offering the service should not jeopardize the realization of other goals and targets. Indeed, the SDGs are linked and interdependent, hence the achievement of one should not compromise the success of the other goal. In the context of broadband, the service should be affordable to all while guaranteeing, least environmental emissions from its deployment and operation.

Several developing economies are yet to meet the targets of the SDGs, especially the Low-Income Countries (LICs). On the other hand, Upper Middle-Income Countries (UMCs) and High-Income Countries (HICs) have made significant steps in addressing SDGs with key milestones in broadband deployment. Surprisingly, few studies have fully investigated the link between universal broadband access, SDGs and environmental sustainability for different economies. Understandably, most broadband research focuses on the cost and technical solutions for providing last mile connectivity. However, there is need to underscore the importance of broadband in realizing SDGs. Documenting and highlighting the role of broadband in realization of the SDGs increases the likelihood of policy makers and analysts in prioritizing and deploying broadband. Therefore, the goal of this paper is to provide a survey and review the research that demonstrate the importance of broadband in supporting SDGs and environmental sustainability. Specifically, the research questions are:

1.  What are the major SDGs linked to broadband connectivity?
2.  What are the key broadband technologies for different target areas?
3.  What are the main challenges and future directions on broadband sustainability research?

Before, focusing on the review, we first provide a quick context of broadband connectivity options in Section II. We then discuss the growing concern of the impact of connectivity options to the environment in section III. In section IV, we discuss our methodology before presenting the results in Section V. In section VI, we discuss the research findings and make recommendations. We then make conclusions in Section VII.

## II.  Broadband Connectivity Options

Several technologies and methods exist for providing broadband to the unconnected. At international level, countries are connected to the global Internet network through undersea optical cables or in some cases via satellites. Once the undersea optical fiber cable lands at the country's coast (landing point), it is routed to the major cities (core nodes) with dense populations. The core nodes are then connected to major towns (regional nodes) from where the final traffic can be delivered to and from the users (last mile). The delivery of the broadband to the final user can be done through various methods such as Fiber-to-the-Home link, wireless (4G, 5G, Wi-Fi) or via satellites. The broadband architecture from the international level to the final user is illustrated in **Fig. 1Error! Reference source not found.**.  However, the choice and viability of the technology is complicated by the demand characteristics of the unconnected population. For instance, a significant portion of the unconnected or underserved population live in rural/remote areas with low population density.

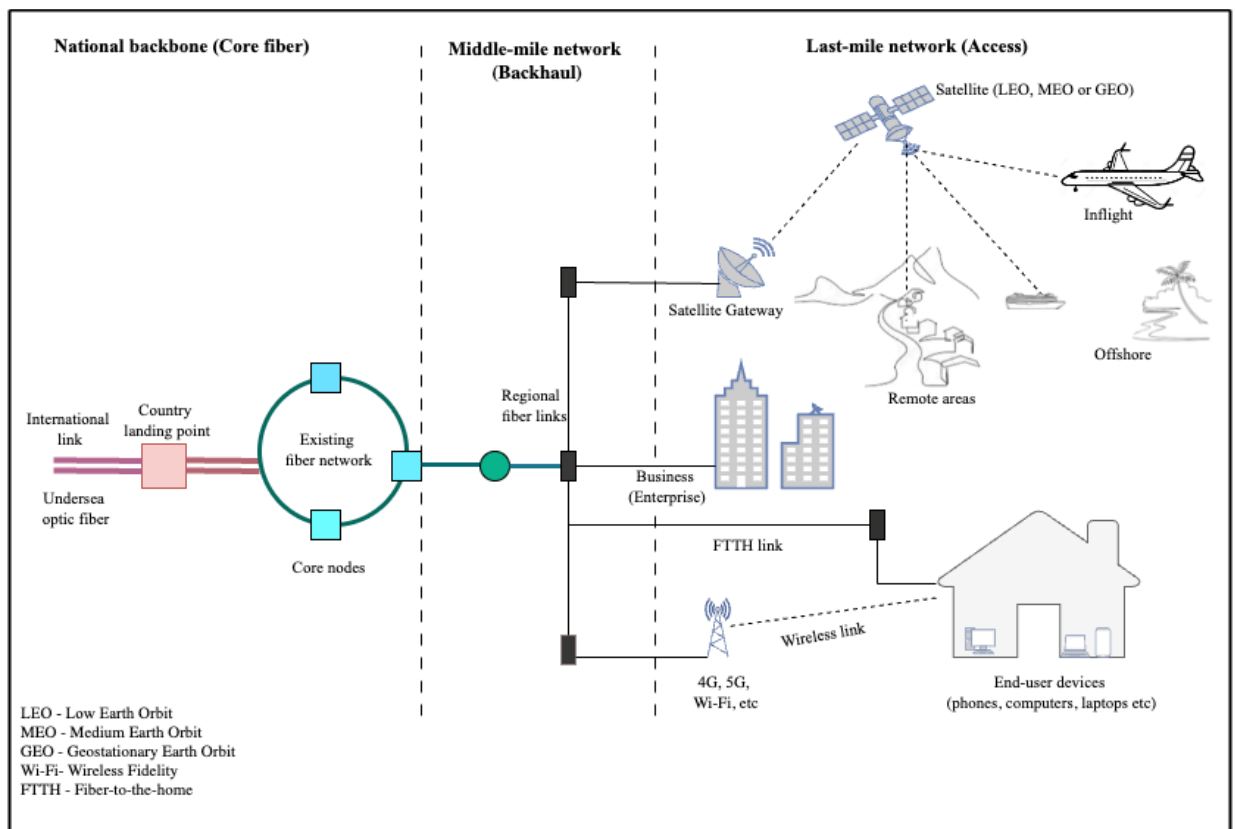

**Fig. 1 | Broadband architecture.** The flow of global Internet traffic to the final user.

Moreover, a bigger percentage of the unconnected population are still living below the poverty line [5]. These challenges complicate the choice of broadband technology to be used as a careful balance must be maintained between quality of service and return on investment. Fixed fiber, Digital Subscriber Line (DSL), fixed wireless access, mobile and satellite broadband are some of the common methods of providing connectivity to unconnected areas. Specifically, mobile broadband has been a dominant method for rural areas due to two reasons. First, majority of the areas have limited legacy infrastructure (roads, electricity, telephone lines) to support other methods such as fiber or DSL. Secondly, there has been a steady increase in smartphone penetration in rural areas, thanks to cheaper prices and alternative payment methods [6]. The main service providers in such cases are Mobile Network Operators (MNOs).

Generally, rural areas are unattractive for Fixed Network Operators (FNOs) due to the high upfront costs vis a vis the projected return on investment. The cost of deploying fiber optic in sparsely populated areas with low Average Revenue Per User (ARPU) does not match their business models. As a result, MNOs have been the major players in the market segment but that is slowly changing. In the last decade, MNOs have been experiencing challenging business times leading to static or declining revenue, with 0.12% decrease in 2022-2023 being an example [7]. The remote and rural areas characterized by low population density and ARPU results into market failure. Despite studies showing that establishment of infrastructure in rural areas can uplift the population from poverty, MNOs and FNOs lack profit motivation to invest in the regions.

Nevertheless, government and international institutions such as UN and ITU have an obligation of providing universal broadband. As a result, emerging technologies such as Low Earth Orbit (LEO) satellites are now fronted as a method of connecting the rural areas. Unlike traditional Geostationary Earth Orbit (GEO) satellites providing broadband, LEO have many advantages. First, the signal/data takes a shorter time to travel from the satellite to the user and back (round trip time), thus can be used in some delay tolerant applications [8]. Secondly, the LEO satellites are designed for shorter life span resulting in lower capital costs that translate to lower service price as compared to GEOs [9]. Moreover, both LEOs and GEOs can provide coverage in remote areas where terrestrial systems fail due to poor topography or inaccessibility such as deserts and mountainous areas.

While providing universal broadband coverage cannot be realized by a single technology, each of the options should be affordable to the user. A single or combination of the three technologies will be vital in solving the connectivity problem. Stakeholders may not agree on the technology choice but the potential of broadband in supporting SDGs is undeniable.

## III.    Broadband and Environmental Sustainability

At the heart of UN's 17 SDGs is the need to protect the environment. In fact, the UN has outlined its commitment to protect the environment in SDG 13 (Climate Action). Research has established that human activities are the major causes of Greenhouse Gas (GHG) emissions resulting to global warming and climate change.  Latest studies shows that the ICT sector accounts for 3.6% of global carbon dioxide ($CO_2$) emissions [10]. Broadband is at the backbone of ICT. Importantly, the broadband infrastructure requires electricity that has a carbon footprint in its generation [11]. Furthermore, the materials such as steel, cement, fiber optic cables used in construction of broadband infrastructure such as antennas, base stations have an associated carbon footprint. Therefore, there is need to ensure that provision of universal broadband coverage does not result in significant GHG emissions that jeopardize the achievement of SDG 13.

The GHG emission concerns from deployment and operation of broadband networks results from the technology options and the increasing demand for high data capacity. Deploying fixed fiber network has been considered as the greenest (1.7 t of $CO_2$ equivalent per user annually) form of broadband [12]. On the other hand, mobile broadband has the highest associated carbon footprint [13]. The high carbon footprint of mobile broadband results from increase in data demand and speeds. New radio systems such as 5G and 6G that seek to provide ultrafast Internet requires thousands of base stations. The base stations run on electricity or diesel that results in higher carbon emissions. For instance, a 2020 study in China revealed that, operation of 5G base stations resulted in 17□5 millions of tons of $CO_2$ equivalent. The legacy mobile systems such as 2G, 3G and 4G also contribute to the emissions due to electricity consumption from the base stations and consumer user equipment. The emissions may be worse in rural/remote areas where offgrid sites are operated using diesel generators.

The issue of GHG emissions has not been a concern for satellite broadband until the emergence of LEO systems. Historically, the small number of satellites launched had negligible emission impacts to the atmosphere [14]. However, the emergence of megaconstellation LEO networks present a new challenge. Several satellites are required in LEO to provide coverage due to two reasons. First, the low altitude (100 – 2000 km) of the satellites results in small coverage area. Secondly, LEO satellites travel between 5-10 km/s hence more are required to guarantee continuity in service due to frequent handovers. Apart from the orbit configuration, the short lifespan (utmost 5 years) of LEO satellites means that constant rocket launches are needed to replace them. Emissions from a single rocket launch may be negligible but servicing larger constellations raises the environmental concerns [15]. Therefore, it is important to evaluate the associated GHG emissions of deploying satellite broadband to achieve balance between meeting the universal coverage mandate and environmental implications.

## IV.   Methodology

Elsevier, IEEE, Springer and Wiley databases were used to search for relevant literature. A search query consisting of keywords was used to ensure that only literature related to broadband connectivity, Internet, SDGs and GHG emissions were selected. No time restriction was applied in the search to ensure that all literature focusing on the role of broadband in supporting SDGs are selected. However, the search was limited to peer reviewed journals and articles. Conference papers and proceedings were excluded from this survey.

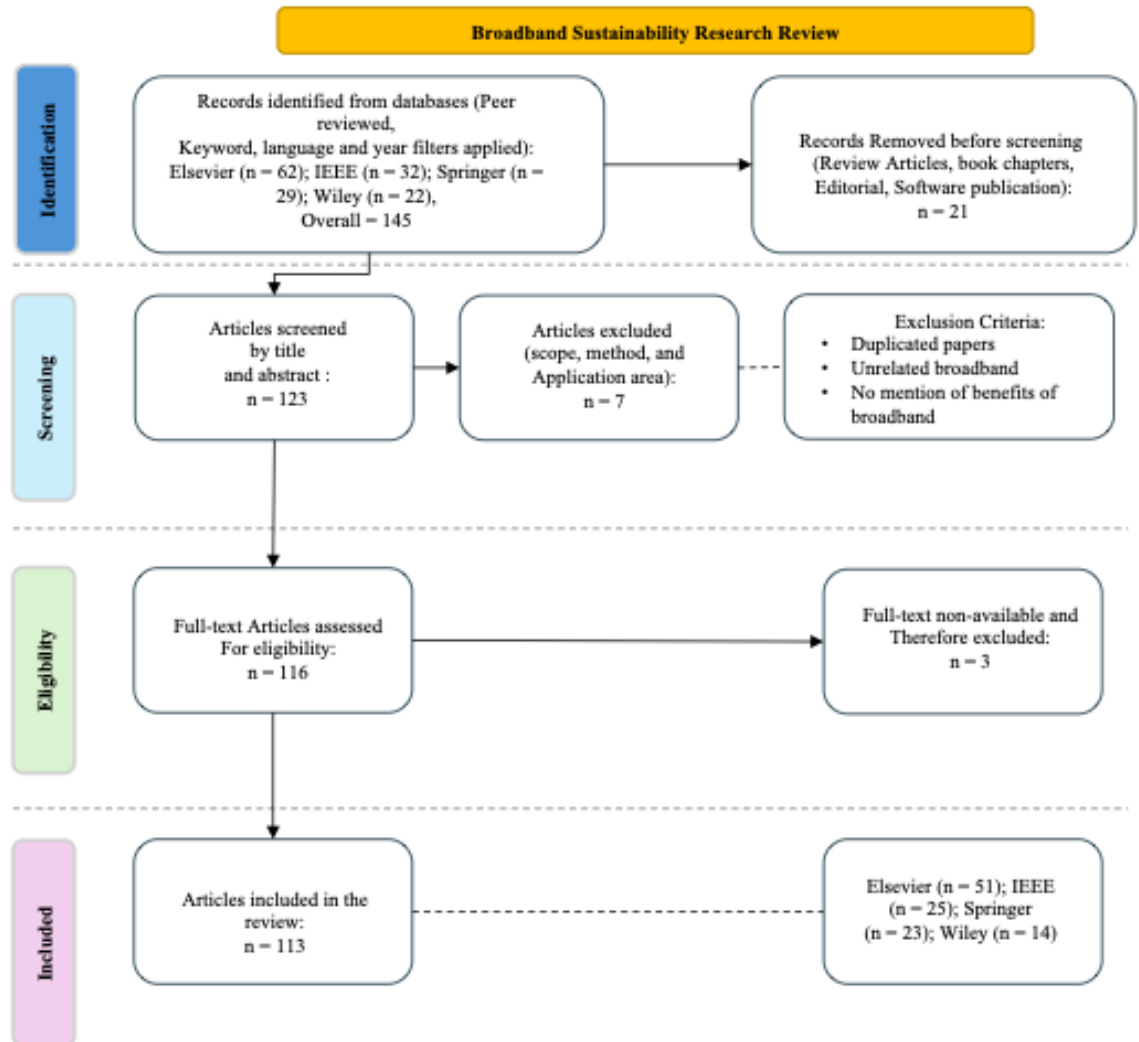

**Fig. 2 | PRISMA flowchart.** PRISMA flowchart followed in reviewing the literature on broadband connectivity, SDGs and sustainability.

The search query yielded a total of 145 literature materials consisting of 62 from Elsevier, 32 from IEEE, 29 from Springer and 22 from Wiley. Next, the evidence-based method for conducting systematic literature review known as Preferred Reporting Items for Systematic Reviews and Meta-Analyses (PRISMA) was adopted [16]. The PRISMA flowchart is illustrated in Fig. 2. Before further screening of the literature, 21 records were removed consisting mainly of book chapters, editorials and software publication resulting to 123 articles.

The screening process entailed removing irrelevant articles based on the scope, method and application area. It specifically involved removing all duplicated papers, those with the words "broadband" in unrelated context and others that mentioned SDGs but not related to Internet connectivity. The screening of the literature resulted into 116 articles. Next, a total of 3 articles were excluded as there was no full-text available resulting in 113 records. Therefore, the 113 articles reviewed included 51 from Elsevier, 25 from IEEE, 23 from Springer and 14 from Wiley.

The 113 articles are processed and analyzed using Python programming language. The pandas python library is specifically used to analyze and group the articles based on the region income and spatial focus, broadband technology, SDG addressed, application area, methodology and author details. Pandas is a python library that provide data structure and functions for manipulation of data [17]. The analytical results are plotted into graphics using R programming language. Interested readers can access the open-source code for reproducibility in the author's GitHub repository account [18].

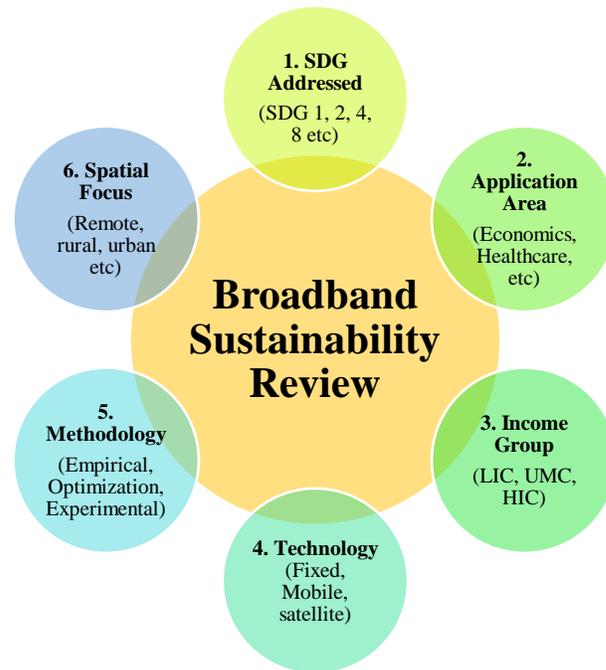

**Fig. 3 | Key areas.** Breakdown of focus areas of the broadband sustainability review.

## V. Results

The results of the analysis of the 113 articles are broken down into six key areas (SDGs addressed, application areas, country income, technology, methodology and spatial focus) as illustrated in Fig. 3 and presented in this section. A full list of all the peer reviewed articles analyzed in this study is provided in Table 1 (A-N) and Table 2 (O-Z).

| Author (s) | Year | Reference | Country Income Group | Broadband Technology | Spatial Focus | SDG Addressed | Methodology |
|---|---|---|---|---|---|---|---|
| Abrardi & Sabatino | 2023 | [19] | HIC | Wired | Urban | SDG 9 | Empirical |
| Aldashev & Batkeyev | 2021 | [20] | UMC | Wired | Rural | SDG 9 | Empirical |
| Anser et al. | 2021 | [21] | UMC | All | All | Not Mentioned | Statistical |
| Bala | 2023 | [22] | HIC | Mobile | All | SDG 9 | Empirical |
| Bhandari et al. | 2024 | [23] | LIC | Mobile | All | Not Mentioned | Experiment & Survey |
| Bolla et al. | 2010 | [24] | HIC | All | All | SDG 13 | Literature Review |
| Briglauer & Gugler | 2019 | [25] | Global | Wired | All | SDG 9 | Statistical |
| Briglauer et al. | 2024 | [26] | Global | Wired | All | Not Mentioned | Literature Review |
| C. Wang & L. Wang | 2024 | [27] | UMC | Wired | Urban | SDG 9 | Empirical |
| Cheng et al. | 2022 | [28] | HIC | Mobile | All | SDG 13 | Optimization & AI |
| Chiha et al. | 2020 | [29] | HIC | Satellite | Rural | SDG 9 | Statistical |
| Clercq et al. | 2023 | [30] | HIC | All | All | Not Mentioned | Empirical |
| Dalwadi et al. | 2023 | [31] | LIC | Mobile | All | Not Mentioned | Literature Review |
| De Souza et al. | 2021 | [32] | UMC | All | Remote | SDG 8 | Empirical |
| del Portillo et al. | 2021 | [33] | HIC | Satellite | Rural | SDG 9 | Optimization & AI |
| Deller & Whitacre | 2019 | [34] | HIC | Wired | Remote | Not Mentioned | Statistical |

| Author (s) | Year | Reference | Country Income Group | Broadband Technology | Spatial Focus | SDG Addressed | Methodology |
|---|---|---|---|---|---|---|---|
| Dhara et al. | 2023 | [35] | LIC | Mobile | All | Not Mentioned | Literature Review |
| Du et al. | 2023 | [36] | UMC | All | Urban | SDG 13 | Empirical |
| Edquist | 2022 | [37] | Global | Mobile | All | SDG 9 | Empirical |
| Edquist et al. | 2018 | [37] | Global | Mobile | All | SDG 8 | Statistical |
| Fadhil et al. | 2023 | [38] | LIC | Fixed Wireless | All | Not Mentioned | Literature Review |
| Ford & Koutsky | 2006 | [39] | HIC | Wired | Rural | SDG 9 | Empirical |
| Gandotra & Jha | 2017 | [40] | Global | Mobile | All | SDG 13 | Literature Review |
| Gandotra et al. | 2017 | [41] | LIC | Mobile | All | Not Mentioned | Optimization & AI |
| Gareeb & Naicker | 2017 | [42] | UMC | All | All | SDG 9 | Statistical |
| Garroussi et al. | 2023 | [43] | HIC | Mobile | All | SDG 13 | Optimization & AI |
| Golard et al. | 2023 | [44] | HIC | Mobile | All | Not Mentioned | Experiment & Survey |
| Golin | 2022 | [45] | HIC | All | All | SDG 3 | Empirical |
| Golin & Romarri | 2024 | [46] | HIC | Wired | Urban | SDG 10 | Empirical |
| Guo et al. | 2022 | [47] | UMC | Mobile | All | SDG 13 | Statistical |
| Gur & Kulesza | 2024 | [48] | HIC | Satellite | Rural | SDG 9 | Empirical |
| Hashemizadeh et al. | 2023 | [49] | Global | All | All | SDG 9 | Empirical |
| He et al. | 2013 | [50] | UMC | Mobile | All | Not Mentioned | Optimization & AI |
| Helmy & Nayak | 2020 | [51] | HIC | Wired | All | Not Mentioned | Optimization & AI |
| Hu et al. | 2023 | [52] | UMC | All | Urban | Not Mentioned | Literature Review |
| Huseien & Shah | 2022 | [53] | Global | Mobile | All | Not Mentioned | Optimization & AI |
| Islam et al. | 2021 | [54] | HIC | Fixed Wireless | Remote | SDG 8 | Experiment & Survey |
| Israr | 2023 | [55] | LIC | Mobile | Urban | SDG 13 | Optimization & AI |
| Jaffer et al. | 2023 | [56] | LIC | Mobile | Urban | Not Mentioned | Optimization & AI |
| Jiang et al. | 2022 | [3] | None | All | Not Applicable | All | Literature Review |
| Jiang et al. | 2024 | [57] | UMC | Wired | Urban | Not Mentioned | Literature Review |
| Joshi et al. | 2024 | [58] | LIC | All | Not Applicable | SDG 4 | Optimization & AI |
| Kandilov & Renkow | 2010 | [59] | HIC | Wired | Rural | SDG 9 | Empirical |
| Karakani | 2024 | [60] | UMC | All | Not Applicable | All | Experiment & Survey |
| Keene et al. | 2024 | [61] | HIC | All | Rural | SDG 10 | Empirical |
| Khan et al. | 2024 | [62] | LIC | Mobile | All | Not Mentioned | Literature Review |
| Kim & Orazem | 2016 | [63] | HIC | Wired | Rural | Not Mentioned | Empirical |
| Kohli et al. | 2024 | [64] | HIC | All | All | SDG 3 | Empirical |
| Kohn | 1997 | [65] | HIC | Satellite | Rural | SDG 9 | Statistical |
| Koutroumpis & Sarri | 2023 | [66] | HIC | Wired | All | Not Mentioned | Empirical |
| Kryszkiewicz et al. | 2022 | [67] | HIC | Mobile | Rural | SDG 9 | Experiment & Survey |
| Li et al. | 2024 | [68] | UMC | All | Urban | Not Mentioned | Empirical |
| Liu et al. | 2014 | [69] | HIC | Mobile | All | Not Mentioned | Statistical |
| Lu et al. | 2016 | [70] | UMC | All | All | SDG 13 | Optimization & AI |
| Ma et al. | 2024 | [71] | UMC | Wired | Urban | SDG 9 | Empirical |
| Ma et al. | 2022 | [72] | UMC | Mobile | All | SDG 9 | Statistical |
| Ma et al. | 2020 | [73] | HIC | All | Rural | SDG 10 | Empirical |
| Maier et al. | 2014 | [2] | HIC | All | All | SDG 9 | Literature Review |
| Mohammed et al. | 2020 | [74] | HIC | Wired | All | Not Mentioned | Statistical |
| Nomikos et al. | 2022 | [75] | HIC | Satellite | All | SDG 9 | Literature Review |

Table 1 Broadband sustainability papers included in the analysis (first author A-N).

| Author (s) | Year | Reference | Country Income Group | Broadband Technology | Spatial Focus | SDG Addressed | Methodology |
|---|---|---|---|---|---|---|---|
| O'Shea et al. | 2023 | [76] | HIC | All | Rural | SDG 3 | Empirical |
| Okundamiya et al. | 2022 | [77] | Global | Mobile | All | SDG 13 | Optimization & AI |
| Orakzai et al. | 2017 | [78] | LIC | Mobile | All | Not Mentioned | Optimization & AI |
| Oruma et al. | 2021 | [79] | LIC | All | Rural | SDG 2 | Literature Review |
| Oughton | 2023 | [80] | Global | Mobile | All | SDG 10 | Empirical |
| Oughton & Russell | 2020 | [81] | HIC | Mobile | Urban | SDG 9 | Experiment & Survey |
| Oughton et al. | 2023 | [82] | HIC | All | All | SDG 9 | Empirical |
| Oughton et al. | 2019 | [83] | HIC | Mobile | Urban | Not Mentioned | Empirical |
| Oughton et al. | 2022 | [84] | Global | Mobile | All | SDG 10 | Empirical |
| Owolabi et al. | 2023 | [85] | LIC | All | All | SDG 9 | Empirical |
| Pan et al. | 2023 | [86] | UMC | All | All | SDG 13 | Empirical |
| Paul et al. | 2023 | [87] | LIC | Fixed Wireless | All | Not Mentioned | Experiment & Survey |
| Peng et al. | 2024 | [88] | UMC | All | All | SDG 13 | Empirical |
| Phillipson et al. | 2013 | [89] | HIC | Wired | Urban | SDG 8 | Empirical |
| Popli et al. | 2021 | [90] | Global | All | All | SDG 13 | Experiment & Survey |
| Renga et al. | 2024 | [91] | HIC | Mobile | All | Not Mentioned | Optimization & AI |
| Robles-Gómez et al. | 2021 | [92] | HIC | All | Urban | SDG 4 | Experiment & Survey |
| Saia | 2023 | [93] | Global | All | All | Not Mentioned | Empirical |
| Saidani et al. | 2024 | [94] | Global | Mobile | All | Not Mentioned | Statistical |
| San Miguel et al. | 2024 | [95] | HIC | All | All | SDG 13 | Empirical |
| Sargam et al. | 2023 | [96] | LIC | Mobile | All | Not Mentioned | Experiment & Survey |
| Savazzi et al. | 2022 | [97] | HIC | Mobile | All | Not Mentioned | Optimization & AI |
| Sboui et al. | 2016 | [98] | HIC | Mobile | All | SDG 13 | Optimization & AI |
| Schaub & Morisi | 2019 | [99] | HIC | All | All | Not Mentioned | Experiment & Survey |
| Schneir et al. | 2019 | [100] | HIC | Mobile | Urban | Not Mentioned | Empirical |
| Shaengchart & Kraiwanit | 2023 | [101] | UMC | Satellite | All | SDG 9 | Experiment & Survey |
| Sharma et al. | 2019 | [102] | LIC | Mobile | All | SDG 9 | Literature Review |
| Shehab et al. | 2021 | [103] | HIC | Mobile | All | SDG 13 | Literature Review |
| Sinclair | 2023 | [104] | HIC | Wired | Urban | Not Mentioned | Experiment & Survey |
| Skinner et al. | 2024 | [105] | HIC | Wired | Rural | SDG 10 | Statistical |
| Srivastava et al. | 2020 | [106] | Global | All | All | SDG 13 | Literature Review |
| Strover et al. | 2024 | [107] | HIC | Wired | Rural | Not Mentioned | Empirical |
| Temesgene et al. | 2017 | [108] | HIC | Mobile | All | SDG 13 | Literature Review |
| Valentín-Sívico et al. | 2023 | [109] | HIC | All | Rural | SDG 9 | Experiment & Survey |
| Wagner & Lee | 2023 | [110] | HIC | Wired | All | SDG 9 | Experiment & Survey |
| Wang & Dong | 2023 | [111] | Global | All | All | SDG 13 | Statistical |
| Wang et el. | 2024 | [112] | UMC | Wired | Rural | SDG 8 | Empirical |
| Wei & Yin | 2024 | [113] | UMC | Wired | Urban | SDG 13 | Empirical |
| Williams & Bergman | 2023 | [114] | HIC | Mobile | All | Not Mentioned | Literature Review |
| Williams et al. | 2022 | [115] | HIC | Mobile | All | SDG 9 | Literature Review |
| Wittig | 2009 | [116] | HIC | Satellite | Rural | SDG 9 | Empirical |
| Wu et al. | 2018 | [117] | None | All | Not Applicable | All | Literature Review |
| Xiang et al. | 2022 | [118] | UMC | Wired | Rural | Not Mentioned | Experiment & Survey |
| Xiao et al. | 2024 | [119] | UMC | All | Urban | SDG 13 | Empirical |
| Xie et al. | 2023 | [120] | Global | All | All | SDG 9 | Literature Review |
| Xu & Zheng | 2023 | [121] | UMC | Wired | Urban | Not Mentioned | Empirical |
| Xu et al. | 2019 | [122] | HIC | Wired | Rural | SDG 10 | Statistical |
| Yang et al. | 2023 | [123] | UMC | All | Urban | Not Mentioned | Empirical |
| Yangming et al. | 2024 | [124] | UMC | All | Rural | SDG 10 | Statistical |
| Zhang | 2021 | [121] | HIC | Wired | All | SDG 9 | Empirical |
| Zhang & Weng | 2024 | [125] | UMC | Wired | Urban | SDG 10 | Empirical |
| Zhang et al. | 2022 | [126] | HIC | All | All | SDG 13 | Literature Review |
| Zhou et al. | 2024 | [127] | UMC | All | Urban | SDG 13 | Empirical |

Table 2 Broadband sustainability papers included in the analysis (first author O-Z).

## a. SDGs Addressed

The delivery of universal broadband is defined by Target 9.c of SDG 9 that states the need to provide universal and affordable Internet access in least developed countries. To this end, several articles were identified that proposed different solutions to meeting SDG 9. A significant portion of the literature concentrated on cheaper and alternative ways of connecting the rural and remote areas with heavy focus on lowering spectrum costs [67]. Some suggested unlicensed spectrum such as community Wi-Fi to deliver the last mile connectivity and guarantee that the local population is connected [2]. Even though SDG 9 is more concerned with providing affordable service in least developed countries, multiple studies investigated the most affordable ways of providing ultra-fast broadband to rural areas in developed countries such as China and the United States [19], [25]. Despite the difference in income group classification of the countries, studies underscored the need to provide modest speed in least developed countries and ultra-fast speed to those with already basic service.

Secondly, the papers addressing SDG 13 focused on the technology and costs of broadband, but a significant amount of authors investigated the environmental cost of broadband infrastructure. While few explicitly mentioned SDG 13 (climate action), the overall objective in the identified broadband papers was to lower the GHG emissions. Some studies quantified the amount of carbon emissions from constructing and operating broadband infrastructure in densely populated areas [36], [88], [113]. Other authors conducted detailed life-cycle assessment of telecommunication infrastructure to determine their contribution to global warming [94]. Most of the articles reviewed linked broadband connectivity to climate change based on the energy requirements of the infrastructure. The link was more direct for ultra-dense networks such as 5G where hundreds of base stations consuming more power are required to meet growing user data demand [55]. As a result, several studies were identified where the authors studied or proposed new hardware and network designs [41], data processing, virtualization, optimization and integration of renewable energy sources in broadband infrastructure to reduce carbon footprint [31], [35], [78], [91]. These studies directly or indirectly address SDG 13 by not only quantifying the carbon footprint of broadband systems but also suggesting ways of lowering it.

In addition to SDG 9 and SDG 13, the survey identified several studies that investigated the role of broadband connectivity in reducing inequalities in societies (SDG 10). The first category of papers identified broadband as a tool of reducing poverty for populations in rural areas. For instance, a study conducted in China demonstrated how broadband access improved quality of elderly care to the rural population [128]. In politics and governance, access to Internet was identified as an enabler to full participation of rural population in democratic process such as voting as well as improved civil education, with Europe [46] and US [99] being a case study. Notably, a bulk of the studies focused on technologies that can provide last mile connectivity to rural and remote areas to ensure that they receive equal service as urban regions. As a result, several authors proposed different satellite configurations [33], [48], pricing strategy and business models [82] to narrow the digital divide prevalent in rural and remote areas.

Furthermore, five studies linked broadband access to SDG 8 in promoting inclusive and sustainable economic growth. In these studies, the authors identified broadband as an enabler and catalyst for innovation, more employment opportunities, business development and economic growth. Some studies conducted in China revealed an increase in employment opportunities after provision and upgrade of broadband speeds in some areas [19], [73]. In other areas, studies revealed a rise in patents and innovations after deployment of faster Internet speeds [25], [129]. The consensus in the studies is that broadband supports innovation and productivity via access to online resources thus enabling value-addition in service provision.

Finally, limited papers mentioned the role of broadband in supporting SDG 2, 3 and 4. Broadband service was identified as a key enabler of precision agriculture that would reduce uncertainty in crop yields, an important step in achieving zero hunger (SDG 2) [79]. Apart from agriculture, broadband connectivity improves the well-being and quality of healthcare (SDG 3). For instance, some studies highlighted the role of Internet in telemedicine to provide affordable mental healthcare service to remote areas during Covid-19 [45], [64], [76]. Lastly, some studies mentioned the role of broadband in improving the quality of education (SDG 4). The studies specifically focused on the role of broadband in meeting the learner's needs in relation to the industry 4.0 demands orchestrated by growth in big data, artificial intelligence, robotics and Internet of Things [58].

## b. Application Area

The highest number of papers reviewed studied the role of broadband in facilitating business and economic growth. The authors related broadband connectivity to business growth, Gross Domestic Product (GDP), employment and innovation. First, the link between broadband access and GDP was common in the papers reviewed especially in developed world [85], [130]. Such studies were more relevant during and after Covid-19 as countries struggled to pivot out of the pandemic. The pandemic period was a reflection point on the role of broadband and the future of digital economy with many studies emerging on the same [112]. Studies showed that broadband access greatly mitigated the socioeconomic impacts of Covid-19 [19]. Notably, pre and post pandemic studies underscored the role of broadband in enhancing employee efficiency and establishment of new businesses in rural areas [71]. For instance, studies focusing in rural South Africa demonstrated an increase in business outlets in rural regions connected to Internet [42], [63], [110]. Furthermore, some survey-based papers recorded an increase in labor productivity, thanks to Internet access at workplace [26], [37].

Secondly, the role of broadband in supporting energy sector to reduce emissions has been extensively studied. A review of the papers revealed the importance of broadband in designing energy efficient systems as well as automating processes to reduce power consumption. Specifically, the papers focused on methods employed by broadband providers to reduce energy consumption of radio access networks while meeting the user's high data demand [77], [91], [115]. On the other hand, some studies emerged on methods of reducing power consumption and prolonging the battery lifespan of mobile user terminals [41]. While the focus of the papers reviewed might be on the provider or end-user, the overall results showed the growing concern of decarbonizing the telecommunication sector.

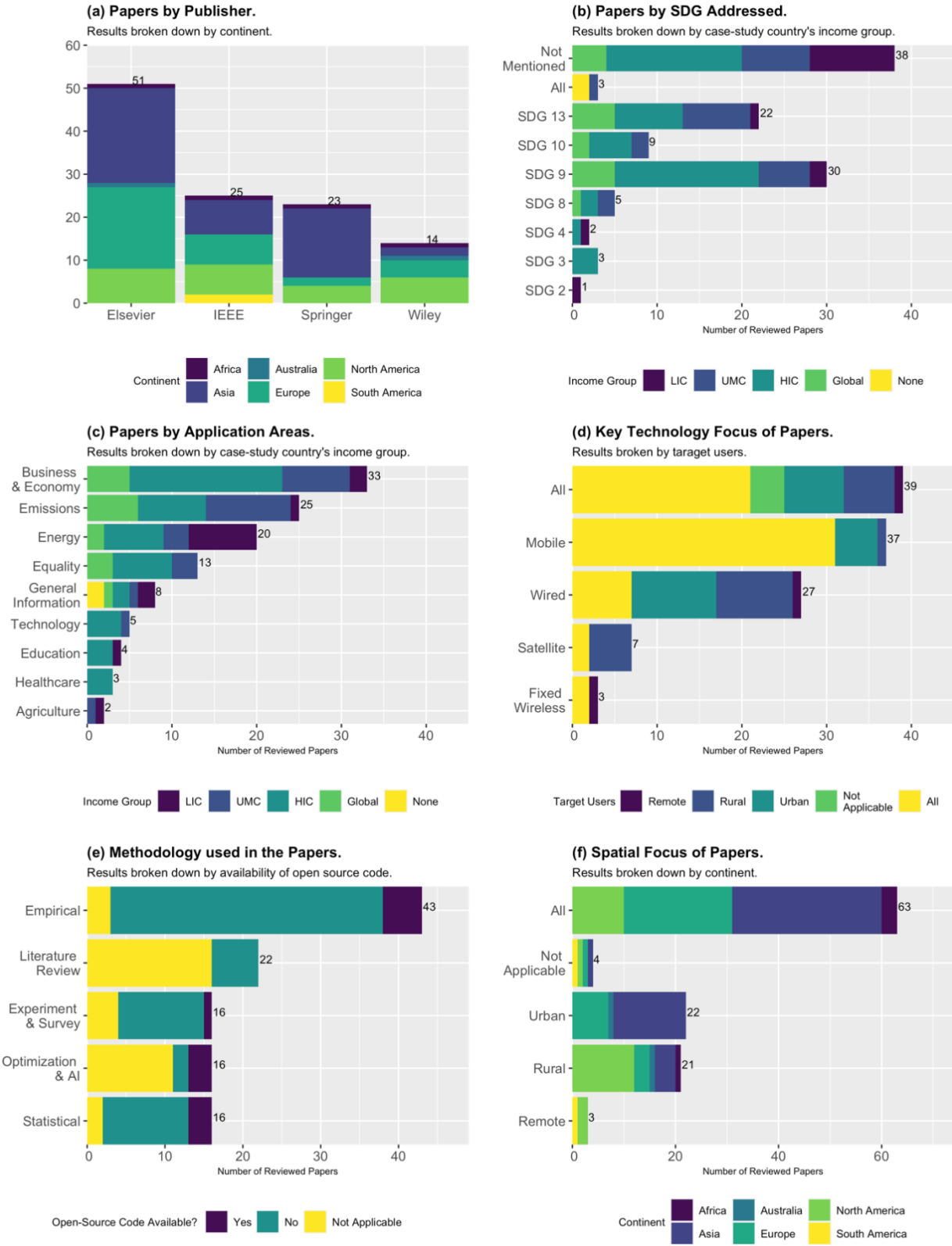

**Fig. 4 | Literature results.** Quantitative literature review results of broadband sustainability.

The need to reduce the carbon footprint of broadband systems drives the interest in investing in greener technology. Indeed, the 2030 projections shows that there will be over 8 billion mobile broadband subscribers [106]. These concerns explain the rise in broadband research focusing on emissions as shown in Fig. 4c. The studies identified in the review are not only limited to broadband connectivity but also extend to other digital technologies that results in carbon emissions [36], [86], [93], [94]. To this end, historical and new empirical data are used across the studies to quantify the carbon footprint of the systems and suggest possible mitigation measures [88], [126]. The results from the samples analyzed in this review are not conclusive but generally indicate that broadband results in carbon emissions but equally reduce it by improving efficiency in other systems and sectors.

Apart from emissions concern, the role of broadband in fostering equality among societies was well documented in 13 articles reviewed in this study. The authors in all the reviewed 13 articles raised the concern on hesitance of broadband operators to provide service in low income households [104]. However, the articles underscored the importance of Internet access in reducing inequality through digital literacy programs, financial inclusion and quality education in rural areas [49], [105], [109]. Moreover, broadband access expand the career and professional development of marginalized communities leading to greater socio-economic opportunities [101]. Availability and uptake of broadband service leads to new opportunities that improve the living standards of societies thus reducing the poverty gap as outlined in the reviewed articles.

Lastly, agriculture, education and healthcare are key areas that broadband technology has impacted positively. Innovative farming practices such as use of IoT sensors require Internet connectivity to provide information such as precipitation, temperature, windspeeds and soil moisture content. Such information are incorporated in precision agriculture models with overall goal of improving yields [79]. In education, the Internet is a platform for delivering online content and facilitating distance learning, hence lowering the education costs [92]. For healthcare, Internet facilitates and supports other services such as telemedicine and drone drug delivery that might otherwise be expensive when done using the conventional methods [22].

## c. Income Group

The SDGs aim to address critical challenges such as poverty, inequality, injustice and climate change by 2030. The aim is to promote prosperity, peace and sustainability especially in least developed countries that are still far behind. As illustrated in the previous section, the reviewed articles have identified the role of broadband in addressing the SDGs. To this end, it is vital to understand the countries where the studies focus on. The hope is that more research is directed in LICs and UMCs compared to HICs.

In Fig. 4b, the addressed SDGs in broadband studies are categorized by income group of the case study country. Most of the studies pertaining to SDG 9, 10 and 13 are common in HICs compared to UMCs and lowest in LICs. These are critical SDGs as SDG 9 aims to improve the state of innovation and infrastructure while SDG 10 focuses on reducing inequalities in societies. SDG 13 on climate action is a global concern as all countries are interested in mitigating climate change.

In terms of application areas, most of the reviewed articles focusing on business and economy are conducted in HICs. Few studies concentrated on LICs with majority of them being case case studies in Sub-Saharan Africa [42], [131]. A similar trend was observed for studies focusing on carbon emissions, energy and equality. No studies were recorded for the role of broadband in healthcare in LICs. On the other hand, the role of broadband in supporting agriculture was evident in studies focusing in LICs and UMCs while absent in HICs.

### d. Broadband Technology

The target area to be served has a significant impact on the choice of broadband technology to be deployed. In the introductory section of this paper, the common broadband technologies were reviewed. Here, we use the categories to report the identified technologies in the papers reviewed and the spatial focus.

First, mobile broadband is the most common technology of choice identified in 37 articles compared to 39 papers that focused on all technologies Fig. 4d. Most studies focused on mobile broadband network systems such as 3G, 4G and 5G with data rates, cost of service provision and carbon footprint as the main thematic areas. Notably, the 5G mobile broadband featured more on carbon emission related works. In terms of spatial focus, most of the mobile broadband technology studies focused on urban markets compared to rural areas. No mobile broadband technology studies were recorded that concentrated on the remote areas.

Secondly, we reviewed articles that focused on wired broadband systems. We defined wired broadband systems as those where the provider and end user devices are connected physically by either fiber optic, copper or any form of cable. Majority of the studies identified, focused on urban and rural areas with a smaller portion studying the deployment of wired networks in remote areas.

Other than wired broadband, 7 articles focused on satellite network systems. The satellite networks identified consisted of GEO, LEO and Medium Earth Orbit (MEO) systems. Even though satellite network systems are aimed at serving the remote areas, none of the reviewed articles focused exclusively on remote areas but in some cases combining all the user areas (remote, rural and urban). Most of the studies focused on rural areas with none using urban areas as a case-study.

Lastly, 3 articles studied the use of fixed wireless technologies in serving users in remote areas. We define fixed wireless to include terrestrial technologies where data is transmitted using radio technologies to stationery users. As a result, the identified studies were those that focused on Wi-Fi and community networks for rural areas with some form of backhaul such as mobile or satellite technology.

### e. Methodology

There are several methodology approaches used to study the relationship between broadband connectivity and SDGs, many of which rely on empirical data and household surveys. Also, optimization, AI and statistical approaches are more evident in studies that investigate measures to reduce energy consumptions and carbon footprint of broadband systems. Lastly, some meta-analysis and systematic literature reviews have been conducted to appraise the role of broadband technology in achieving some SDGs. In this review, we focus on empirical, literature review, experimental/survey, optimization/AI and statistical methodologies used in studying the role of broadband in addressing SDGs.

We define empirical methodology as a study where some existing data is used to study the correlation between broadband and SDGs. In 43 reviewed articles, authors used different datasets including but not limited to census, employment, number of school enrollment, number of registered businesses and broadband providers in an area to study the relationship. The use of empirical data was more evident in papers that investigated the relationship between broadband and economic development [25], [85], [130]. The results provided numerical conclusions explaining the role of broadband in addressing related SDGs such as SDG 9. In some cases, school attendance [105] and business employment data [26], [37], [66] were used to explain the impact of broadband in improving the standards of living in rural areas.

Other than empirical papers, 22 of the 113 reviewed articles were literature review or some form of meta-analysis focusing on the role of broadband in sustainable development. Most of the identified reviews explained how broadband can be used to leverage and support other sectors including agriculture [131], education [58] and healthcare [64]. Other authors researched on how developing countries could leverage internet to narrow the digital divide while also reaping on the cascading effect of broadband on other sectors [60]. Several works focused on new radio systems such as 5G and 6G entailing the opportunities and challenges they present [55], [108], [114], [132], [133]. Most of the works identified in the 5/6G category delved into cellular planning and alternative methods of powering the new radio systems to reduce their energy consumption and carbon footprint.

As a result, 16 articles applying optimization and AI techniques in suggesting energy efficient broadband networks were identified. Optimization methodology involves finding the best solution out of existing alternatives for a constrained problem. The reviewed studies focused on finding the best solution that minimized the power consumption of the hardware systems while serving maximum users [41], [51], [97]. To this end, some papers focused on multicell user planning in an area [50]. Others focused on AI methods such as reinforcement learning to dynamically manage constrained spectrum and energy sources especially in satellite systems to serve maximum users [134], [135], [136]. Furthermore, the multi-objective optimization techniques were more popular in studies that link the capacity, coverage, cost and carbon footprint of broadband systems. For instance, some authors applied linear integer programming in lowering the power consumption of 5G cellular networks [28], [56], [77], [91] while one study used the approach to maximize satellite coverage while lowering per user monthly costs [33].

In 16 reviewed papers, experimental and survey methodology was used to either study the cause-effect relationship or model the broadband systems. The survey approach was especially common for case-study papers seeking to evaluate the impact of broadband deployment on government's programs in rural areas. For instance, a survey was conducted in New Zealand [118] and Australia [104] to evaluate broadband usage and impact of government subsidy programs on rural users. Likewise, experimental approaches was popular in studies focusing on measuring the power consumption and carbon footprint of broadband hardware systems [44], [56]. Additionally, some studies experimented on new hardware such as antenna designs to measure the capacity [87] while others evaluated the service quality of Internet speed against the costs available to rural areas [81], [96], [101], [109].

Irrespective of the methodology used, 16 studies applied statistical techniques to add rigor to the outcome. Thus, statistical approaches such as point pattern and regression analysis were evident in some studies especially those that correlated broadband and SDGs [105], [122], [124]. The correlation studies also related demand and supply side broadband factors and their interplay in proposing the best solution for narrowing digital divide [34]. The use of advanced statistical techniques thus provided further evidence on various hypothesis tested in the papers to underscore the role of broadband in development and need to close the digital divide.

## f. Spatial Focus

Lastly, the spatial focus of the research work is vital in addressing digital divide as part of SDGs. The least developed areas are the key focus of broadband policymakers to bring the associated population online. A significant portion of the unconnected population lives in continents dominated by countries classified as LICs. However, the literature review results shows that majority of the papers analyzed focused on connecting the rural and remote population in North America that is dominated by HICs compared to UMC and LIC. The number of papers focusing on rural and remote population in continents dominated by LICs such as Africa and South America are low. Notably, 63 papers focused on all population areas with majority concentrating in Asia and Europe. Fig. 5 shows the breakdown of the country of focus based on the income group and the target area of the broadband research.

**First author's country and research focus breakdown.**
Country income classification and targeted user's area investigated

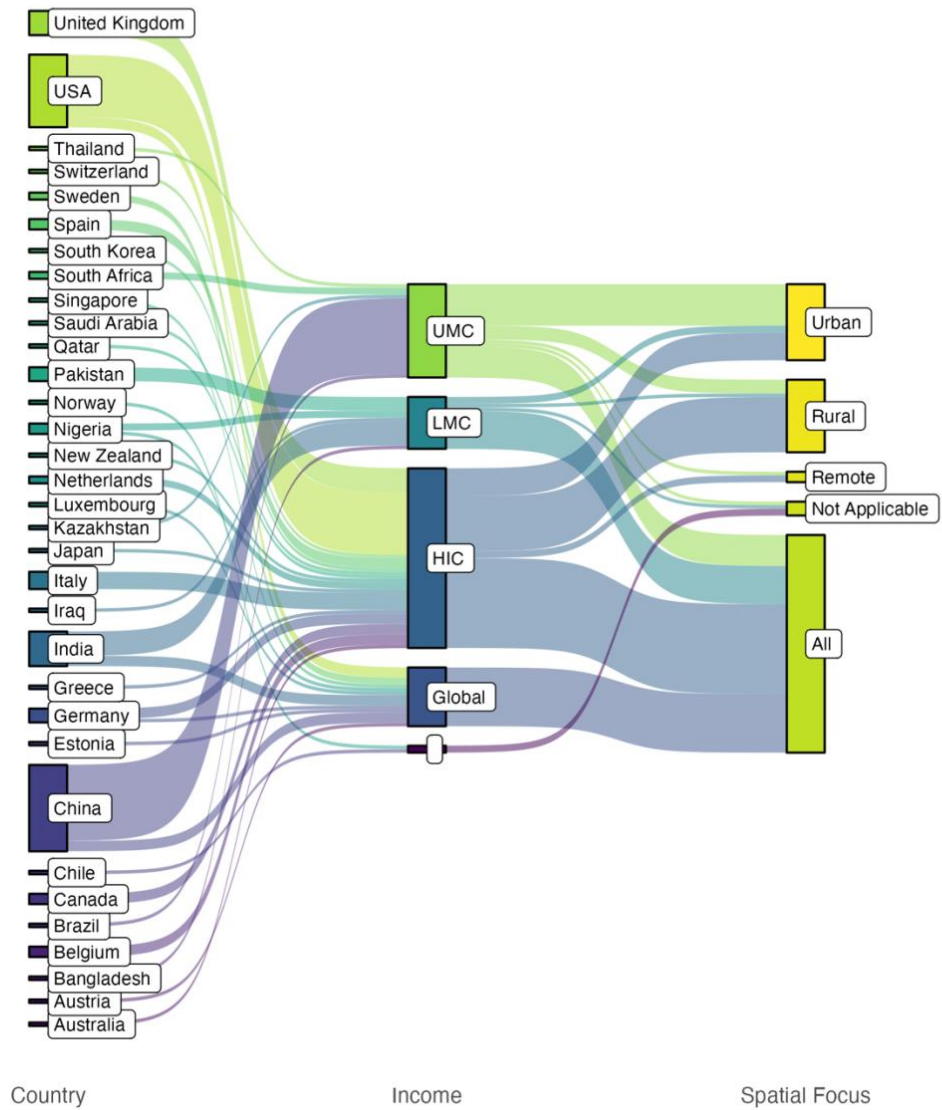

**Fig. 5 | Country breakdown.** Country of research focus grouped by income category and the target user area.

## Geospatial Sample Composition of the Reviewed Articles

Now that we have analyzed the articles based on the SDGs addressed, application area, country income group, broadband technology, methodology and spatial focus, we now present the composition of all the papers in the survey. We also present the distribution of the location of first author as well as the total paper output per country, Fig. 6. This helps to appreciate the general role of broadband in supporting development and interest in achievement of SDGs by the first-approaching 2030 deadline.

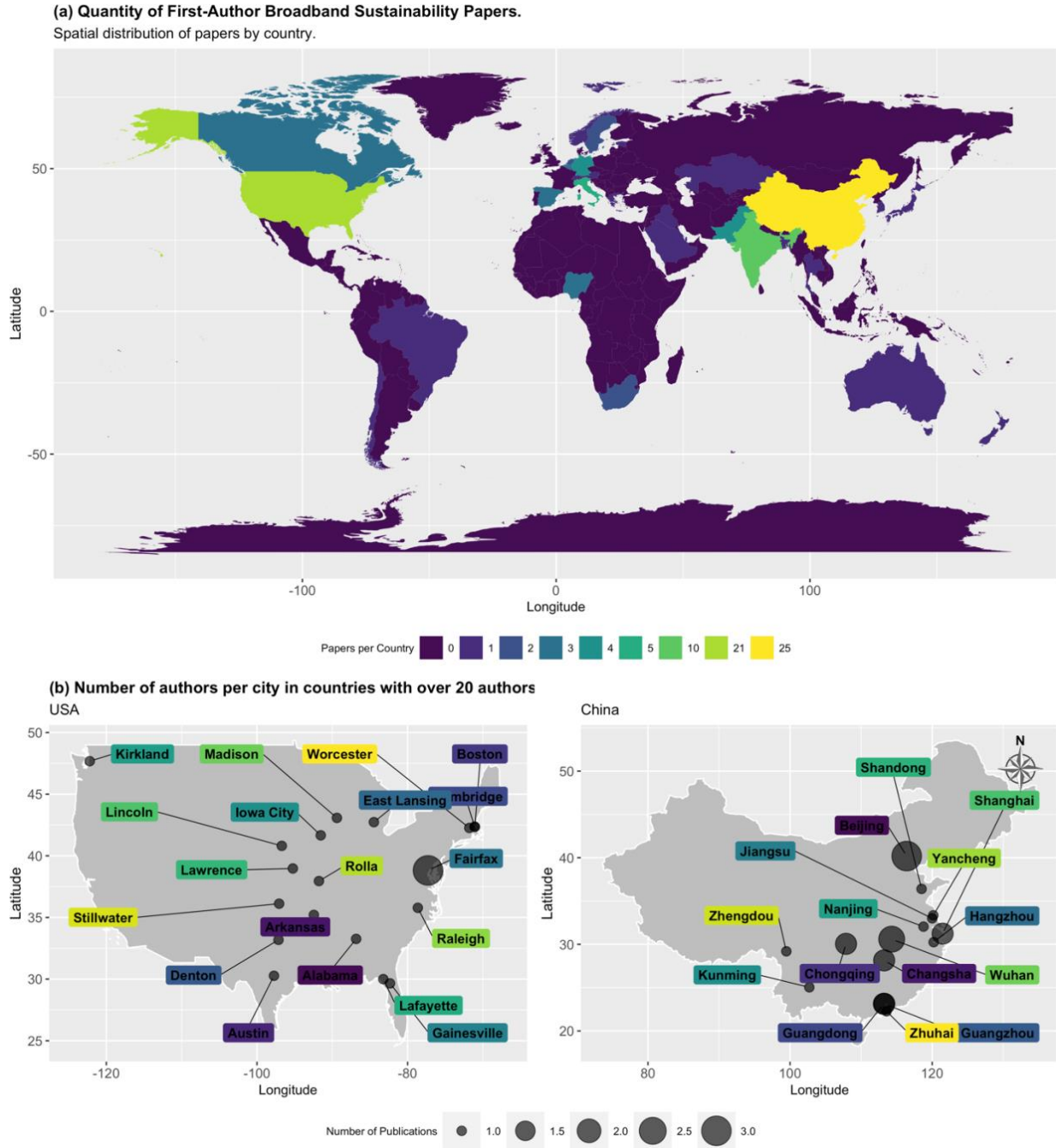

**Fig. 6 | Paper Distribution.** Location of broadband sustainability sample papers first-author.

Firstly, the number of broadband sustainability papers per database broken down per continent of the first author is presented in Fig. 4a. The Elsevier database has the highest number of papers (51) compared to IEEE (25), Springer (23) and Wiley with the least number, 14. The high number of publications by Elsevier is justified since it has several journals that focus on technology policy and impacts on society such as Telecommunication Policy and Technological Forecasting and Social Change journals. On the other hand, IEEE has a significant number of journals, thanks to its contribution to telecommunication engineering research. Majority of the works focusing on technology to connect the rural and remote areas are published on IEEE hence the second largest number of papers. In terms of continents, Asia, Europe and North America have the highest number of publications. They are then followed by modest contributions from Africa, South America and Australia.

Secondly, Fig. 6a shows that broadband sustainability research is not popular in most countries with majority registering just a single paper. Many of the countries have not recorded instances of a first author paper on broadband sustainability research. This is particularly troubling in South America and Africa, because these are the continents with highest number of developing countries that needs to address SDGs as well as the largest portion of unconnected broadband users. Contrastingly, the United States (Fig. 6b) and China records the highest number of first author broadband sustainability papers at 21 and 25 respectively. Other than the population of the two countries, their interest in broadband technology, funding priorities and the impact on the people and economy might be the driving factor.

The United States' federal government has been at the forefront of connecting the underserved communities. The rationale is that broadband access is a powerful tool of providing equal access to socioeconomic opportunities. As a result, most of the first author paper publications are recorded in the Midwest and Southern cities and States of Arkansas, Alabama, Missouri, Nebraska and Iowa. For China, most of the publications are concentrated in the Eastern major cities consisting of Beijing, Shanghai, Wuhan and Guangzhou. The high frequency of papers in these major cities might be attributed to the growing concern of the need to decarbonize the telecommunication sector. As a result, 18% of the reviewed papers focused on the energy consumption and the carbon footprint of broadband systems. Thus, China being highest contributor of broadband research, a significant portion of the papers might have focused on the sustainability as opposed to just narrowing the digital divide.

## VI.   Discussion

Having extensively reviewed the literature on broadband sustainability focusing on the contextual and technical aspects of the papers, we now synthesize our findings in the context of narrowing the digital divide while addressing the SDGs. To do this, we revisit the three research questions outlined in the introductory section as follows.

### a.   What are the major SDGs linked to broadband connectivity?

Our results showed that even though the analyzed papers related broadband connectivity to development, majority do not explicitly mention or connect specific SDGs to the technology. However, a significant number of papers mentioned SDG 9 in relation to providing universal and affordable broadband access (Target 9.c). This is especially common in HICs which is logical given that countries such as USA and Canada have been at the forefront of connecting their rural and remote populations. In most of the papers focusing on HICs, the objective was connecting the remote population using innovative methods as well as funding the projects through government subsidies, incentives and partnerships.

Secondly, SDG 13 on climate action featured extensively on the analyzed papers. The papers addressing this SDG concentrated on ways of reducing power consumption of the broadband networks to reduce their carbon footprint. Indeed, decarbonizing the telecommunication sector has become an emerging concern given that the industry contributes over 3.6% carbon dioxide ($CO_2$) emissions globally [10]. This is further exacerbated by the growing number of broadband users and the emergence of new ultra dense radio technologies (5G, 6G and LEO satellite systems). To this end, most of the research papers were published in China and India (UMCs) and HICs (USA, Italy, UK). The number of papers around the climate action agenda underscores the importance of balancing technology deployment and environmental sustainability.

Across the seven identified SDGs, there was a reasonable quantity of research focusing on SDG 10 specifically Target 10.3. Target 10.3 demands that countries should reduce inequalities by eliminating any discriminatory laws, policies and practices to empower everyone socially, economically and politically. The papers addressing SDG 10 focused on ways of eliminating barriers to broadband deployment in rural and remote areas. Some of the identified practices were the need to reduce or eliminate spectrum fees, obligation to provide service in remote areas and government subsidies. Other technical works focused on broadband infrastructure planning and configuration to reduce the cost of deployment and operation while guaranteeing modest speeds. The analyzed papers recognized the potential of broadband in reducing inequalities in rural/remote areas through creation of employment opportunities, improving quality of education and promoting businesses.

Notably, the focus of papers conducted in LICs concentrated on the role of broadband in supporting basic services such as education and agriculture. For instance, broadband was mentioned as a vital tool for facilitating online learning especially in resource-constrained schools and education centers. Similarly, broadband was identified as an essential supporting technology for new farming practices such as precision agriculture. The studies illustrated the need for LICs in regions such as Sub-Saharan Africa to appreciate the role of broadband beyond just connecting the population but also supporting other sectors of the economy to address SDGs by 2030.

To sum up, the research on role of broadband in supporting SDGs vary according to the economic classification of the country. The top priority of research conducted in HICs and UMCs focused more on technical aspects of broadband and the need to mitigate climate change. The research focus is justified given that such countries have moved beyond major coverage issues to the need for ultra-fast and power consuming networks to meet the growing data demand. On the other hand, the research in LICs were domiciled on reducing inequality and leveraging Internet to support agriculture, education and healthcare.

### b. What are the key broadband technologies for different target areas?

This review identified and grouped broadband technologies into mobile, wired, satellite and fixed wireless. These are the available technology options for serving users in urban, rural and remote areas. By far, mobile and wired systems were the dominant broadband technology in serving users especially in urban areas. This is logical as the high population density and ARPU justify the investment in urban areas. The revenue generated by the users is sufficient to keep the MNOs and FNOs operating within their profit margins. On the other hand, these technologies are unviable for rural and remote areas as identified in the papers analyzed. The implications are that such areas will remain unconnected unless government partners, incentivize or mandate the operators to provide obligatory service in the regions. Coincidentally, the rural and remote areas require broadband connectivity to support other development sectors including agriculture, business, education and healthcare to address the goals and targets of SDGs by 2030.

Notably, the SDGs seek to realize social, economic and political prosperity of the developing world by 2030. To this end, rural and remote areas should be connected to broadband service to access similar opportunities as their urban counterparts. Unfortunately, our results shows that continents with the highest number of developing nations such as Africa and South America have the least number of research focusing on rural and remote population. This can be attributed to different research priorities or lack of research funding to support such endeavors. In contrast, developed nations (UMCs and HICs) are interested in using technologies such as internet to addressing the emerging challenges as growing data demand, power consumption and resultant emissions. The danger in the divergent broadband research is the likelihood of applying or implementing business models that might have worked in UMCs and HICs to LICs leading to failure and further digital divide. As a result, slow progress will be realized in the journey to connect the unconnected and address the SDGs by 2030.

## c. What are the main challenges and future directions on broadband sustainability research?

Having discussed the implication of the literature review results, we now embark on the contribution of this work on broadband sustainability. Several challenges must be overcome to narrow the digital divide and fast-track the achievement of the SDGs by 2030 especially in developing economies. To this end, we propose four recommendations in broadband deployment as follows.

### 1. Open access to data and global models.

The results of the review revealed that most of the broadband sustainability research is conducted in economies in HICs and UMCs compared to LIC. However, this is a problem as a significant portion of the unconnected population that are in urgent need to address the SDGs are found in LICs. That does not imply that the research conducted in HICs cannot be replicated or applied to address the problem of digital divide. Indeed, the demand and supply metrics of broadband is usually similar in economies with major difference on the level of target user's disposable income, existing infrastructure, perceived relevance and level of government funding. These modeling inputs can be customized on a country-by-country basis provided the methodology is generic.

Therefore, we recommend that future research should prioritize reproducibility and open access to data, and assumptions that underpin their methodologies. Setting the work and the methodology in a generic way enables users from any country to customize and tailor the research to conduct case study for individual target areas. Indeed, some studies [80], [82], [84] focusing on open source datasets and modeling approach have emerged to assist policymakers and operators across the globe in broadband, planning, deployment and viability assessment. Such tools are characterized by access to the data, codes, assumptions and results that users in LICs can replicate without necessarily committing extra resources at research and planning level.

2. Embrace multi-technology approach in last mile connectivity.

Using broadband to support SDGs requires economies to narrow the digital divide. Several technologies exist at operators and policymaker's disposal. However, each of the technologies have different shortcomings. For instance, MNOs are unable to provide services without incurring losses in remote and rural areas. The topography and settlement patterns of the users in remote and rural areas especially in LICs increases the infrastructure deployment cost of wired systems such as fiber. On the other hand, satellites can provide coverage, but the initial cost of user terminals and monthly subscriptions limits the potential of using them in connecting the remote/rural population.  Furthermore, wired and satellite technologies rely on availability of other infrastructure systems such as electricity and roads.

To this end, we recommend that studies focusing on the role of broadband in SDGs should explore the integration of all the technologies. Narrowing the digital divide will require a blend of two or more technologies. For instance, satellites could be used as backhaul for fixed wireless access networks provided through community and business partnerships. Such approaches will not only avail Internet to the remote/rural population but also provide new employment and business opportunities. Fortunately, 39 of the 113 papers analyzed in this review studied the relationship between multiple broadband technologies and development.

3. More research is required on emerging technologies.

As per the second recommendation, narrowing the digital divide requires the integration of different technologies with no single one capable of solving the problem. There has been a resurgence of new technologies especially in mobile broadband including 5G and 6G. The interest in these technologies is driven by the need to build systems that meet the high data demand requirements. At the same, LEO satellite broadband technology has achieved big milestones with most of the industries in testing while others in deployment and operation phase [137]. However, most of the papers analyzed in this review focused more on 5G and 6G mobile broadband technologies compared to LEO satellite systems. Moreover, the 5/6G papers were biased towards the technical requirements and configurations as well as the issue of power consumption to reduce carbon footprint instead of the costs. Truly, more research on sustainability of broadband is necessary but equally bringing the offline population online is also an urgent need.

As a result, we recommend that new broadband research should focus more on the business and financial model of LEO systems instead of concentrating on just the technical engineering aspects. Indeed, some studies have already began evaluating the viability of LEO satellites in addressing the Internet needs of rural/remote areas [29], [33], [138] while others are also focusing on their environmental impact [139]. Focusing on capacity, coverage, cost and emissions of LEO systems will provide more detailed information that operators and policymakers can rely on to assess their viability in narrowing the digital divide and addressing SDGs in the developing economies.

4. Research works should apply Integrated methods focusing on technical, business and broadband policies.

As the need to close the digital divide grows, detailed research is required to make data-driven decisions. Currently, most of the papers analyzed in this review applied siloed approach where engineering and cost assessment of broadband technology are modeled distinctly. Most of the papers focused exclusively on the technology, cost or policies. A negligible amount of work has been done on integrated models but fall short of directly connecting broadband technology to SDGs [81], [83], [140]. However, it is important to conduct integrated detailed modeling of broadband networks as operators and policymakers subject any project to cost benefit analysis. Even in the case of best technology, a breakdown of business model will lead to failure in deployment of an infrastructure. Moreover, using the best technology and implementing a viable business model requires better policies.

To this end, we suggest that studies focusing on broadband sustainability should implement integrated modeling approaches encompassing the technology, cost and policy framework. Such studies allow the consumers to test different technologies, business and financial models against various policies to evaluate their viability. For instance, the data rate and cost of deploying 5G in an area can be subjected under various policies such as full or passive infrastructure sharing to evaluate the profitability as done in some previous studies [141], [142]. Similarly, conventionally expensive broadband options such as satellite can be tested against different import tariff, incentives, quotas and spectrum allocation requirements to determine their affordability. Developing such integrated models will improve the quality of broadband sustainability work available to research community and countries struggling to fast-track the achievement of SDGs by 2030.

## VII.   Conclusion

The role of broadband connectivity in spurring development has been studied since the rollout of Internet. With the 2030 deadline fast approaching and the worry that SDGs might not be realized by then, there is need to highlight, explore and analyze the role of Internet in fast-tracking their achievements. To this end, we identified 145 papers with 113 peer reviewed journals selected for this detailed meta-review. The selected papers were reviewed based on 6 dimensions consisting of SDGs addressed, application areas, broadband technologies, methodology of study, income classification of the country and the spatial focus. Our review unearthed interesting trends and challenges in broadband sustainability research.

For example, despite Africa and South America continents having the highest number of developing countries that need to fast-track SDGs, they recorded the least number of broadband sustainability studies. On the other hand, developed countries such as China and the USA recorded the highest number of broadband sustainability research focusing on SDG 13 (climate action) and SDG 9 (industry, innovation and infrastructure). Importantly, there is a surge in the number of papers focusing on the carbon footprint of broadband systems, specifically 5G.

Importantly, we climaxed our review with four key recommendations pertaining to the need to narrow the digital divide and broadband as a tool for supporting SDGs. The recommendations reflect on the need for transparency in data and research methodology to facilitate reproducibility, multi-technology solution, focus on emerging technologies, and integrated modeling approach. The hope is that following these proposals will increase interest in broadband sustainability research as well as suggest viable options in the struggle to narrow the digital divide and fast-track the realization of SDGs.

To finalize, we identified contextual and methodological limitation. Contextually, some of the studies identified did not explicitly mentioned SDGs but investigated the role of broadband connectivity in addressing the thematic areas, targets and indicators of the SDGs. Methodologically, our search of the papers was only limited to peer reviewed journal articles published in English. Consequently, papers published in other languages such as French, Mandarin, Indian, Arabic were excluded hence underrepresenting the contribution of some countries. Lastly, the study was limited to peer reviewed journals, thus conference papers with valid results were omitted. Future literature review may benefit from authors in the other languages as well as including reputable peer-reviewed conference papers.

## Acknowledgment

The authors would like to thank the Department of Geography and Geoinformation Science at George Mason University for funding this work.